# Spin-state ordering and magnetic structures in YBaCo$_2$O$_{5.5/5.44}$

D.D. Khalyavin[*]

*Institute of Solid State and Semiconductors Physics, National Academy of Sciences, P. Brovka str. 17, 220072 Minsk, Belarus*

D.N. Argyriou and U. Amann

*Hahn-Meitner-Institut, Glienicker Straße 100, Berlin D-14109, Germany*

A.A. Yaremchenko and V.V. Kharton

*Department of Ceramics and Glass Engineering, CICECO, University of Aveiro, 3810-193 Aveiro, Portugal*

The antiferromagnetic - ferromagnetic phase transition in YBaCo$_2$O$_{5.50}$ and YBaCo$_2$O$_{5.44}$ cobaltites with different types of oxygen-ion ordering in the [YO$_{0.5/0.44}$] layers has been studied by neutron powder diffraction. Using the magnetic symmetry arguments in combination with group-theoretical analysis, the crystal and magnetic structures above and below the phase transformation temperature, T$_i$, were determined and successfully refined. In both cases, the proposed models involve a spin-state ordering between diamagnetic ($t_{2g}^6 e_g^0$, $S = 0$) and paramagnetic ($t_{2g}^4 e_g^2$, $S = 2$) Co$^{3+}$ ions with octahedral coordination. Electronic ordering results in a nonzero spontaneous magnetic moment in the high-temperature magnetic phases with isotropic negative exchange interactions. In the case of YBaCo$_2$O$_{5.5}$, the phase transformation does not change *Pmma* ($2a_p \times 2a_p \times 2a_p$) symmetry of the crystal structure. The wave vectors of magnetic structures above and below T$_i$ are ***k*** = 0 and ***k*** = ***c***$^*$/2, respectively. In the case of YBaCo$_2$O$_{5.44}$, a crossover *P4/nmm* ($3\sqrt{2}a_p \times 3\sqrt{2}a_p \times 2a_p$) → *I4/mmm* ($3\sqrt{2}a_p \times 3\sqrt{2}a_p \times 4a_p$) was involved to solve the low-temperature magnetic structure. The wave vectors in both high-temperature and low-temperature magnetic phases are ***k*** = 0. Mechanisms of the phase transformation in both compositions are discussed in the light of obtained magnetic structures. The proposed spin configurations were compared with other models reported in literature.

[*]**To whom correspondence should be addressed:** khalyav@ifttp.bas-net.by

**Introduction**

Cobalt-containing oxide compounds attract a great attention in recent years due to a variety of magnetic and electrical properties related to the spin, charge and orbital degrees of freedom. One additional factor, expanding the number of possible magnetic ground states and enriching the magnetic phase diagrams, is related to the electronic configuration (spin-state) of cobalt cations. A phenomenon of thermally activated spin-state crossover is well known in $LaCoO_3$.[1,2] The different electronic configurations of $Co^{3+}$ appear as a result of the competition between the interatomic exchange interaction ($J$) and the crystal field splitting ($\Delta$). In a purely ionic model, two electronic configurations for octahedrally coordinated $Co^{3+}$ ions are possible: high-spin ($t_{2g}^4 e_g^2$, $S = 2$; at $3J>\Delta$) and low-spin ($t_{2g}^6 e_g^0$, $S = 0$; at $3J<\Delta$). A strong hybridisation of the Co $3d$ orbitals with the oxygen $2p$ orbitals was assumed to stabilise the intermediate-spin state ($t_{2g}^5 e_g^1$, $S = 1$) due to formation of the $d^6 + d^7\underline{L}$ mixed configuration.[3] The intermediate-spin state was concluded to be the first excitation in $LaCoO_3$ from the low-temperature diamagnetic ground state. However, very recent inelastic neutron scattering[4] and electron spin resonance experiments[5,6], and also theoretical calculations based on the unrestricted Hartree-Fock method[7] and the generalized gradient approximation[8], all indicate that the high-spin triplet is the first exited state for $Co^{3+}$ ions in this perovskite. Moreover, the intermediate-spin state which is expected to be the ground state for the square pyramidal configuration because of the crystal field symmetry removing the degeneracy between $d_{3r^2-z^2}$ and $d_{3x^2-y^2}$ orbitals, was not observed either in $Sr_2CoO_3Cl$ oxychloride[9,10] or in $LnBaCo_2O_{5+\delta}$ (Ln-lanthanide or Y),[11-16] both containing $Co^{3+}$ ions in the fivefold pyramidal coordination. These findings cast doubt the ability of oxygen-coordinated $Co^{3+}$ ions to adopt the intermediate-spin configuration.

The latter example is a family of cobaltites isostructural to $YBaCuFeO_{5+\delta}$,[17,18] where the crystal lattice can be represented as a sequence of $[CoO_2]$-$[BaO]$-$[CoO_2]$-$[LnO_\delta]$ layers stacked along the *c* axis. When the concentration of oxygen vacancies in the $[LnO_\delta]$ layers is maximum ($\delta=0$), all cobalt ions are located within square pyramids. As $\delta$ increases, extra oxygen ions occupy vacant sites in $[LnO_\delta]$ layer, thus providing an octahedral environment for a part of the Co ions. Three types of superstructure, $3a_p \times 3a_p \times 2a_p$ (332) at $\delta=0.44$,[18-22] $a_p \times 2a_p \times 2a_p$ (122) at $\delta=0.50$[17,18] and $2a_p \times 2a_p \times 2a_p$ (222) at $\delta=0.75$[23-25], may be formed due to ordering of the extra oxygen (Fig. 1); these correspond to tetragonal (*P4/mmm*), orthorhombic (*Pmmm*), and again tetragonal (*P4/mmm*) symmetry of the crystal structure, respectively. Formation of the 122 superstructure was observed for Ln = Pr-Ho or Y,[18,26] whereas the 222 and 332 superstructures are only stable for large (Pr-Gd)[23-25] and small (Tb, Dy, Ho, Y)[18-22] lanthanide cations, correspondingly.

Metamagnetic behavior conjugated with giant magnetoresistance phenomenon is known for the compounds of 122 type.[17,18,27,28] This structure motif involves penta- and hexacoordinated $Co^{3+}$ ions in the ratio 1:1. A ferromagnetic component, which appears above $T_i \sim 250$ K and can be induced by external magnetic field below this temperature, is an important point at issue. The most common point of view relates to the model presuming equal concentrations of the low-spin and ferromagneticaly ordered intermediate-spin cations.[29-38] The different electronic configurations are supposed to be adopted by $Co^{3+}$ coordinated by oxygen in different ways. In spite of a successful explanation of many physical properties of $LnBaCo_2O_{5.5}$, this model has however serious inconsistencies, particularly with respect to the neutron diffraction data.[39-43] The uncertainties discussed above raise additional doubts.

In previous work,[44] a model involving a chess-board like ordering between the high- and low-spin Co ions in the octahedral positions and uniform high-spin configuration of the pyramidally coordinated cobalt, was proposed. This model, based on the symmetry consideration and analysis of the known experimental data, predicts a strong negative isotropic superexchange via the $e_g$ - $p_\sigma$ - $e_g$ type coupling between the paramagnetic Co cations. The spontaneous magnetization originates from the high-spin octahedral sublattice.[44] This model was extrapolated to explain the magnetic properties of materials with the 332 type superstructure demonstrating similar antiferromagnetic - ferromagnetic transition near $T_i \sim 180$ K. The present work is focused on neutron diffraction studies of the cobaltites with both the 122 and 332 structural types in a vicinity of the phase transition at $T_i$, where the spontaneous magnetization takes place. $YBaCo_2O_{5+\delta}$ was chosen as a model composition taking into account the possibility to stabilize both types of superstructure. A special attention is centered on the nature of the ferromagnetic component. The different models proposed in literature are compared and analyzed on the basis of the obtained neutron diffraction data and symmetry arguments. The magnetic structures related to the spin-state ordering phenomena, proposed in previous work,[44] were found to be in excellent agreement with the neutron diffraction experiment and to satisfy symmetry constrains.

**Experimental**

$YBaCo_2O_{5+\delta}$ samples with $\delta$ close to 0.50 and 0.44 were prepared by a solid-state reaction method using conditions similar to those described by Akahoshi and Ueda.[21] The starting materials, namely $Y_2O_3$ (99.99% Aldrich), $BaCO_3$ (99%, Merck) and $Co_3O_4$ (99%, Aldrich), were mixed in an agate mortar and reacted at 1373 K in pure oxygen gas for 10 h. The firing procedure with intermediate regrindings was repeated several times to remove impurity phases. Finally, the powder was cooled from 1373 K to room temperature at the rate of 50 K/h. X-ray diffraction (XRD) revealed that the obtained material has an orthorhombic symmetry specific for the 122 type superstructure. For the preparation of cobaltite with the 322 crystal structure, the 122-sample was annealed in oxygen at 580 K for 5 h with subsequent rapid cooling down to room temperature.

The oxygen content in these samples was determined by thermogravimetric analysis (TGA) performed using a Setaram SetSys 16/18 instrument. The TGA procedure included heating (5 K/min) to 1223 K in a flow of dry air, equilibration with air at 1223 K for 2 h, flushing the apparatus with argon during 1 h, and then reduction in flowing 10%$H_2$-90%$N_2$ mixture at 1223-1323 K for 13 h. One example of the TGA curve on reduction is given in the insert of Figure 2. For the samples annealed in oxygen at 1373 K with slow cooling and at 580 K with quenching, the values of δ at room temperature were 0.52±0.02 and 0.46±0.02, respectively (Fig.2). For comparison, analogous TGA procedure was also performed for an $YBaCo_2O_{5+\delta}$ sample equilibrated at atmospheric oxygen pressure during annealing at 1373 K and slow cooling (2 K/min) in air; the calculated value of δ was 0.38±0.02. Note that, unlike $PrBaCo_2O_{5+\delta}$,[25] no maximum in the δ(T) curve was observed for the $YBaCo_2O_{5+\delta}$ sample equilibrated in air (Fig. 2).

The dc magnetization was measured using a SQUID magnetometer (Quantum Design, MPMS-5) in the temperature range 4.2 - 300 K. The curves registered at 10 kOe after cooling in zero field (ZFC) clearly demonstrate the phase transitions at $T_i$ ~ 260 K and 160 K for the samples with the oxygen content 5.52 and 5.46, respectively (Fig. 3). The change of the magnetic susceptibility below $T_L$~ 190 K for the sample with the 122 superstructure can be associated with an additional magnetic phase transition similar to that reported by Plakhty et al.[41] and Baran et al.[45] for $TbBaCo_2O_{5.5}$ ($T_L$ ~ 170 K).

Neutron diffraction experiments were carried out in the Berlin Neutron Scattering Center (BENSC) at the Hahn-Meithner-Institute. Two powder diffractometers were used, namely the fine-resolution powder diffractometer (E9) and the high-intensity flat-cone- and powder diffractometer (E2). The former, with incident neutrons of wavelength λ = 1.7974 Å and resolution $\Delta d/d$ ~ $2 \cdot 10^{-3}$, was used for the crystal structure refinement. The second diffractometer (λ = 2.39 Å), equipped with a position sensitive detector providing a very high intensity, is efficient for the magnetic structure investigations. The obtained data were refined by the Rietveld method using the Fullprof suite.[46] Group-theoretical calculations were performed with the aid of the ISOTROPY software.[47]

**Results and discussion**

**YBaCo$_2$O$_{5.5}$.** In correlation with the magnetization data (Fig. 3), additional reflections in the neutron diffraction patterns appear below T$_C$ ~ 290 K. In the temperature range 260 K < T < 290 K where the spontaneous magnetisation exists, the magnetic reflections can be indexed with the wave vector, $\boldsymbol{k} = \boldsymbol{a}^*/2$, applied to the $a_p \times 2a_p \times 2a_p$ unit cell characteristic of the 122 crystal structure (Fig. 4). The strongest magnetic reflection (1/2,1,1) at the $d$-spacing ~ 4.43 Å is usually associated with the $\boldsymbol{G}$-type antifferomagnetic structure, where each Co ion is antiferromagnetically coupled with its nearest neighbours. This observation rules out a ferromagnetic nature of the spontaneous moment, and is consistent with the results reported by Fouth et al.[40] for NdBaCo$_2$O$_{5.47}$, by Frontera et al.[43] for PrBaCo$_2$O$_{5.5}$ and by Plakhty et al.[31] for TbBaCo$_2$O$_{5.54}$. One can conclude, therefore, that in all cases, the magnetic structures and the origin of the spontaneous magnetisation are very similar. A careful examination of the neutron diffraction data of YBaCo$_2$O$_{5.5}$ revealed also that there is a ferromagnetic contribution in some of the nuclear reflections such as, for instance, (0,1,0).

As Co ions occupy two independent positions (2$r$ octahedral and 2$q$ pyramidal) in the *P4/mmm* ($a_p \times 2a_p \times 2a_p$) space group, the following possibilities should be considered: (i) the spin configurations in the octahedral and pyramidal positions are characterized by the two different wave vectors ($\boldsymbol{k} = \boldsymbol{a}^*/2$ and $\boldsymbol{k} = 0$); and (ii) there is a single wave vector for the both sublattices. The first possibility is unlikely since it is impossible to avoid frustrations in the magnetic structure in the case of two distinct wave vectors for the positions with identical transformational properties. Indeed, attempts to refine the neutron diffraction data in the models involving basis functions of both the $\boldsymbol{k} = \boldsymbol{a}^*/2$ and $\boldsymbol{k} = 0$ wave vector groups failed. Assuming the single $\boldsymbol{k} = \boldsymbol{a}^*/2$ wave vector is incompatible with the ferromagnetic component which should disappear at the antitranslation along the $\boldsymbol{a}$ axis. This means that the crystal structure of YBaCo$_2$O$_{5.5}$ is more complex than shown in Figure 1b, and should adopt the $2a_p \times 2a_p \times 2a_p$ unit cell. In latter case, the magnetic structure is characterised by the $\boldsymbol{k} = 0$ wave vector, and the symmetry condition for a nonzero spontaneous magnetic moment is satisfied. The phase transition with the $\boldsymbol{k} = \boldsymbol{a}^*/2$ modulation vector is considered to occur due to an electronic ordering in the cobalt sublattice. Applying the group-theoretical methods,[48-51] all isotropy subgroups of *Pmmm* space group were found to be associated with the X ($\boldsymbol{k} = \boldsymbol{a}^*/2$) point. Inspection of the subgroups revealed that only two of them, *Pmmm* and *Pmma*, split the 2$q$ and 2$r$ cobalt positions in the parent *Pmmm* space group. These subgroups are associated with the $X_1^+$ and $X_2^-$ irreducible representations, respectively.[52] The difference between them relates to the location of equivalent pairs of Co ions in the $2a_p \times 2a_p \times 2a_p$ unit cell. In the case of isomorphic phase transition, Co ions in two pairs of non-equivalent pyramidal (2*q*, 2*s*) and octahedral (2*r*, 2*t*) positions are linked by a mirror plane perpendicular to the ***c*** axis. In the second case, Co cations in the corresponding positions (2*e*, 2*e* and 2*f*, 2*f*, respectively) are linked by the glide plane with translational component along the ***a*** axis and, therefore, these are shifted in a half-period along this direction. The basis functions for irreducible representations of the wave vector group in the case of *Pmmm* and *Pmma* symmetry are presented in Refs. 41 and 44, respectively. In both cases, the magnetic representations on the cobalt positions consist of six one-dimensional irreducible representations of the paramagnetic space group symmetry. These results indicate that an orthogonal ferromagnetic component is forbidden by symmetry and a ferrimagnetic component should be responsible for the spontaneous magnetisation. However, due to the different distribution of equivalent cobalt sites in the unit cell of these structures, different magnetic modes should produce the ferrimagnetic component. To classify the magnetic modes, two vectors ***F***$_\alpha$ and ***A***$_\alpha$ can be introduced using the basis functions ***S***$_i$ (*i* numerates cobalt in equivalent positions) from Refs. 41, 44:

$$\boldsymbol{F}_\alpha = \boldsymbol{S}_1 + \boldsymbol{S}_2 \quad \boldsymbol{A}_\alpha = \boldsymbol{S}_1 - \boldsymbol{S}_2 \ (\alpha = q, r, s, t \text{ for } Pmmm \text{ and } e, e, f, f \text{ for } Pmma) \qquad (1)$$

Any magnetic mode can be represented as a linear combination of these vectors taken over all the independent cobalt positions in the unit cell. Obviously, only the configurations involving ***F***$_\alpha$ vectors can produce a nonzero spontaneous moment. As mentioned above, the analysis of magnetic reflections suggests the ***G*** - type spin configuration. This magnetic mode for the *Pmma* and *Pmmm* symmetry can be presented as:

$$\boldsymbol{G}(Pmma) = \boldsymbol{F}_e - \boldsymbol{F}_e + \boldsymbol{F}_f - \boldsymbol{F}_f \qquad \boldsymbol{G}(Pmmm) = \boldsymbol{A}_q - \boldsymbol{A}_s + \boldsymbol{A}_r - \boldsymbol{A}_t \qquad (2)$$

Thus, a nonzero magnetic moment in the unit cell may only be expected in the first case. This result indicates that the real symmetry of the crystal structure of YBaCo$_2$O$_{5.5}$ is *Pmma* with the 2$a_p$×2$a_p$×2$a_p$ unit cell dimension. One should mention that such conclusion would be very complicated without the magnetic symmetry arguments. Indeed, a careful inspection of the neutron diffraction data did not reveal any superstructure reflections directly confirming the *Pmma* space group. However, these reflections were observed by Chernenkov et al. in a single crystal X-ray diffraction experiment on the GdBaCo$_2$O$_{5.5}$ composition.[53] In the present work, the powder neutron diffraction patterns were simulated using atomic displacements reported in Ref.53; it was found that a visualisation of these reflections requires unreasonably high statistic parameters and, apparently, can never be reached in powder diffraction experiments. In spite of this fact, the crystal structure parameters for YBaCo$_2$O$_{5.5}$ were refined in the *Pmma* space group (Fig. 5). The results (Table 1) were then used to refine the magnetic structure at T= 275 K, using the data recorded on the E2 diffractometer (Fig. 6) and allowing different values of the magnetic moments for the different cobalt positions. The best fitting ($R_{nuc}$ = 3.31 %, $R_{mag}$ = 8.79 %) was obtained with the magnetic moments directed along the *a* axis and the values 1.3(1) $\mu_B$, 1.3(1) $\mu_B$ and 0.9(1) $\mu_B$, 0.0(1) $\mu_B$ for the cobalt cations occupying positions with the same coordination, respectively. The results clearly demonstrate that the electronic ordering suggested above is a spin-state ordering between the diamagnetic and paramagnetic cobalt ions appearing in the pyramidal or octahedral positions of the parent 122 structure.[54] Considering the refinement results, these two possibilities are indistinguishable. However, the octahedral position is assumed more preferable for the superstructure development since: (i) the sixfold coordinated Co$^{3+}$ ions it are widely accepted to adopt the low-spin configuration in LnCoO$_3$ perovskites,[55] (ii) the high-spin state was confirmed both experimentally[9-16] and theoretically[10] to be a ground state for the square-pyramidal coordination of Co$^{3+}$, (iii) many physical properties of LnBaCo$_2$O$_{5.5}$ can be explained on the basis of this assumption and the model can be successfully extrapolated to the 332 structural type.[44] Thus, according to the obtained magnetic structure (Fig. 7a), the spontaneous magnetic moment originating mainly from the octahedral position is equal to 0.23 $\mu_B$ per Co ion, which is in a reasonable agreement with the magnetisation data.

The most similar result relates to the model proposed by Plakhty et al. for TbBaCo$_2$O$_{5.54}$.[41] The authors used the group-theoretical arguments to analyze neutron diffraction data and the obtained spin structures satisfy the symmetry constraints.[41] According to Plakhty et al., the ferrimagnetic component comes from the pyramidally-coordinated Co positions in the 2$a_p$×2$a_p$×2$a_p$ unit cell (*Pmma* space group), and none of the four positions has magnetic moment close to zero. The conclusion[41] that the ferrimagnetic component originates mainly from two non-equivalent pyramidal sites was directly obtained from the refinement, which is not unambiguous. In the present work the magnetic structure solution within the same ferrimagnetic *G* - type mode was found strongly dependent on the starting parameters in the refinement. For instance, the magnetic structure was refined in two stages. At the first stage, the magnetic moments in the pyramidal positions were constrained to be equal to each other, whereas the moments in octahedral sites were varied by different parameters and one of the positions was initially fixed at zero. Independently of the starting magnetic-moment values for the nonzero positions, a convergence was reached with the moments 1.3(1) $\mu_B$ for pyramids and 0.8(1) $\mu_B$ / 0.1(1) $\mu_B$ for the two octahedral positions. At the second stage, all the four positions were independent. Then, the model by Plakhty et al. was applied as starting in the refinement of the magnetic structure. The convergence was reached with good reliability factors ($R_{nuc}$ = 3.31 %, $R_{mag}$ = 8.82 %) and the values of the magnetic moments 1.8(1) $\mu_B$, 0.8(1) $\mu_B$ and 0.5(1) $\mu_B$, 0.5(1) $\mu_B$ for pyramidal and octahedral cobalt positions, respectively. Both these models provide a very accurate description of the magnetic intensity. Therefore, only additional arguments based on the common sense and the ability of model to explain the variety of observed physical properties of $LnBaCo_2O_{5.5}$ compositions can distinguish between these two solutions. The model with zero magnetic moment in one of the position (preferably octahedral) provides a reasonable and clear picture for the electronic type ordering and physical properties of these materials.[44]

An original non-collinear model, based on a single crystal neutron diffraction experiment, was proposed by Soda et al.[42,56] The model operates with *Pmmm* symmetry ($a_p \times 2a_p \times 2a_p$ unit cell) of the crystal structure and assumes $Co^{3+}$ ions in the octahedral position to be in the low-spin state. Magnetic moments in the pyramidal position have two nonzero components, ferromagnetic along the ***a*** axis and antiferromagnetic along the ***b*** axes of the *Pmmm* space group. As discussed in Ref. 44, this model contradicts the symmetry constraints as the pyramidal position involves basis functions of more than a single irreducible representation. Since there are no any additional symmetry arguments explaining why the concept of a single irreducible representation can be broken in this case (except for an accidental degeneracy), this model is considered as unlikely. Moreover, testing of this model in the refinement gave much worse agreement ($R_{nuc}$ = 3.64 %, $R_{mag}$ = 43.4 %) in comparison with the Plakty et al.[41] and our models.

Below $T_i \sim$ 265 K, in accordance with the magnetisation measurements, a rearrangement of the spin structure occurs (Fig. 4). Reflections of the new antiferromagnetic phase can be indexed by doubling the $2a_p \times 2a_p \times 2a_p$ unit cell along the ***c*** axis (wave vector ***k*** = ***c****/2). Following Ref. 44, a model for the refinement was obtained from the previous one by introducing anti-translation along the ***c*** direction (Fig. 7b). The crystal structure parameters were obtained by fitting of the data recorded on the E9 diffracotomer (Table 1). Then, these parameters were fixed to refine the magnetic structure using the E2 diffraction data. Both patterns were collected at 250 K. Again, the refinement was done in two steps, initially by assuming equal values for the pyramids and one zero octahedral position, and then without any constraints. A description of the magnetic intensities was accurate with good reliability factors, $R_{nuc}$ = 4.19 %, $R_{mag}$ = 10.1 % (Fig. 8). The obtained values of the magnetic moments for the octahedral and pyramidal positions are 2.1(1) $\mu_B$, 0.1(1) $\mu_B$ and 1.6(1) $\mu_B$, 1.3(1) $\mu_B$, respectively. These are far from saturation, making it difficult to directly conclude on the cobalt spin-state. However, the substantially high value for one of the octahedral positions (2.1(1) $\mu_B$) and a negative exchange of these cations with pyramidal sublattice are both consistent with the high-spin electronic configuration for all paramagnetic cobalt ions.[57] This model naturally comes from the previous magnetic structure and provides a clear mechanism for the magnetic phase transition.[44] The driving force is a competition between the weak positive exchange interactions via oxygen-deficient [YO$_{0.5}$] layers and the frustrations due to thermal excitations of magnetic moment on the diamagnetic cobalt ions. In this spin configuration, an exited cation can be antiferromagnetically coupled only with three octahedrally coordinated neighbours and ferromagnetically with the fourth one (Fig. 7d). The transition to the state with spontaneous magnetisation enables to avoid frustrations since all nearest neighbors of the exited cation have antiparallel spin in this phase (Fig. 7c).

Similar agreement with the experiment data can be achieved using the spin-state ordered model proposed by Fauth et al.[40] Therefore, only the arguments based on logical interconnection with the high-temperature magnetic phase and a successful explanation of many physical properties of LnBaCo$_2$O$_{5.5}$ testify the validity of the magnetic structure presented here.[44]

The spin configuration suggested by Plakhty et al.[41] is very similar to our model. Despite the fact that the model[41] does not involve a cobalt position with the magnetic moment very close to zero, the moment of one Co position is much lower than that of others, 0.28(8) $\mu_B$. This situation can be explained by the presence of a small amount of misplaced pyramids due to imperfect oxygen ordering in the [TbO$_{0.5}$] layers, as suggested by Frontera et al.[43] for PrBaCo$_2$O$_{5.5}$. Hence, the spin structures proposed in this work and in Ref. 41 are almost identical.

The antiferromagnetic structure discussed above is stable down to $T_L$ ~ 190 K. At lower temperatures, a new magnetic phase is formed; the phase transition is similar to that reported by Plakhty et al.[41] for TbBaCo$_2$O$_{5.54}$ at $T_L$ ~ 170 K. The neutron diffraction patterns in both cases are also similar (Fig. 4). For the new phase, all magnetic reflections characteristic of the previous antiferromagnetic structure are present; in addition, the (1/2,1,1) reflection appears again. Based on the extinction law of the magnetic reflections, Plakhty et al.[41] concluded that the crystal lattice should transform from *Pmma* ($2a_p \times 2a_p \times 2a_p$) into *Pcca* ($2a_p \times 2a_p \times 4a_p$); the new magnetic structure was constructed using the basis functions of irreducible representations of the *Pcca* space group. In the present work, an attempt was made to refine our data using the latter model, no satisfactory results were obtained. The magnetic order arguments used by Plakhty et al.[41] lead to conclusion that the unit cell diminution (translational symmetry) should be changed at the phase transition temperature from $2a_p \times 2a_p \times 2a_p$ to $2a_p \times 2a_p \times 4a_p$; the *Pcca* space group was chosen as the highest subgroup of *Pmma* with the modulation vector $\boldsymbol{k} = \boldsymbol{c}^*/2$. The authors[41] supposed that the superstructure appears due to ordering of Co$^{3+}$ spin/orbital states. However, this space group keeps all four independent cobalt positions with the same arrangement (chessboard type) in the unit cell as for high-temperature *Pmma*. The difference appears in splitting of the 4$h$ Tb position into two independent 4$c$ ones. Thus, one has to conclude that a new type of electronic ordering in the cobalt sublattice cannot be the reason of the symmetry change. The consideration of the isotropy subgroups of the parent *Pmmm* ($a_p \times 2a_p \times 2a_p$) space group, associated with the U ($\boldsymbol{k} = \boldsymbol{a}^*/2 + \boldsymbol{c}^*/2$) point, seems to be more appropriate approach for the crystal and magnetic structures solution. Only the space groups splitting the 2$q$ and 2$r$ initial cobalt positions should be taken into account for the calculations of magnetic representations and the corresponding basis functions. This analysis is in progress now, but since this magnetic structure does not relate directly to the magnetic phase transition at $T_i$, this problem will be discussed separately in the case of its successful solution.

**YBaCo$_2$O$_{5.44}$.** As literature information on the refinement of this structural type is very scarce, a brief introductory discussion on the crystal structure is necessary. According to Zhou[19,20] and Akahoshi and Ueda,[21] YBaCo$_2$O$_{5+\delta}$ (0.25<$\delta$≤0.44) perovskites exhibit a tetragonal $3a_p \times 3a_p \times 2a_p$ type superstructure. An ordering of extra oxygen ions in [YO$_\delta$] layers was assumed to be the origin of the primitive unit cell tripling. Possible models were proposed, without a quantitative structure refinement. Later, tetragonal symmetry (*P4/mmm*) with the same unit cell dimension was suggested for TbBaCo$_{1.88}$Fe$_{0.12}$O$_{5+\delta}$ on the basis of neutron powder diffraction experiment.[22] The most adequate structural model (Fig. 9a) was determined and some structural parameters were refined.[22] However, a low resolution did not allow to establish atomic positions precisely. In the present work, the model[22] was used as a starting one to refine the obtained high-resolution neutron powder diffraction data. The refinement showed two structure solutions with very close reliability factors. The difference between these models relates to the Y3 and O8 displacements from their highly symmetric positions (Fig. 9b, c).

It is difficult to distinguish between these two solutions if only considering the reliability factors. However, in the first case (Fig. 9b) the refinement converges with an unreasonably high degree of oxygen disordering between the O8(4$k$) and O9(4$o$) sites; the latter site should be entirely vacant in a perfectly ordered 332 crystal structure. The second model (Fig. 9c) gives quite reasonable occupancies of these oxygen positions and was thus considered preferable. A careful examination of the room-temperature diffraction data did not reveal any additional reflections proving that the actual symmetry is lower than *P4/mmm*. As for YBaCo$_2$O$_{5.5}$, however, the magnetic symmetry arguments require a more complex crystal structure for the successful solution of magnetic structure. Determination of the crystal lattice requires to consider magnetic contribution to the neutron diffraction patterns below $T_C \sim 290$ K.

According to the magnetization measurements (Fig. 3), $YBaCo_2O_{5.44}$ exhibits a spontaneous moment in the temperature range 160 K<T<290 K, comparable with that observed for $YBaCo_2O_{5.5}$. In this temperature range, additional magnetic reflections can be indexed with the wave vector $\mathbf{k} = \mathbf{a}^*/2 + \mathbf{b}^*/2$. Again, the strongest magnetic reflection (3/2,3/2,1) at the $d$-spacing ~ 4.42 Å indicates the $\mathbf{G}$ type antiferromagnetic spin configuration (Fig. 10). Furthermore, a magnetic contribution into several nuclear reflections, e.g. (1,0,0), is clearly observed. By analogy with $YBaCo_2O_{5.5}$, an electronic ordering in the octahedral position of the parent $P4/mmm$ ($3a_p \times 3a_p \times 2a_p$) structure (Fig. 9a) was assumed in order to account for the ferromagnetic component and to satisfy the $\mathbf{k} = 0$ symmetry condition. To identify the actual symmetry of $YBaCo_2O_{5.44}$, it is necessary to consider isotropy subgroups of the $P4/mmm$ space group, associated with the M ($\mathbf{k} = \mathbf{a}^*/2 + \mathbf{b}^*/2$) point. Only the maximal subgroups, which split the 8r octahedral position in the parent structure, are taken into consideration. This requirement is satisfied by four subgroups, namely $P4/mmm$, $P4/mmm$, $P4/nmm$ and $P4/nmm$ produced by the $M_1^+$, $M_4^+$, $M_2^-$ and $M_3^-$ irreducible representations, respectively. The arrangement of the non-equivalent cobalt positions in these subgroups, in all cases related to the parent one as:

$$\mathbf{a}_s = \mathbf{a}_p + \mathbf{b}_p; \quad \mathbf{b}_s = \mathbf{b}_p - \mathbf{a}_p; \quad \mathbf{c}_s = \mathbf{c}_p \qquad (3)$$

($\mathbf{a}_s$, $\mathbf{b}_s$, $\mathbf{c}_s$ and $\mathbf{a}_p$, $\mathbf{b}_p$, $\mathbf{c}_p$ being the lattice vectors of the subgroup and parent group, respectively), is shown in Figure 11. For each octahedral cobalt site, the positive and negative directions of spin in the $\mathbf{G}$ - type configuration are also presented. One can clearly see that this magnetic configuration may only produce a non-compensated magnetic moment for the $P4/nmm$ subgroup coresponding to the $M_2^-$ irreducible representation. Thus, at this stage, the actual symmetry of $YBaCo_2O_{5.44}$ is assumed to be $P4/nmm$ ($3\sqrt{2}a_p \times 3\sqrt{2}a_p \times 2a_p$), Figure 11a. This space group was used to refine the high-resolution neutron diffraction data recorded at the room temperature and at 200 K. The starting model was generated by the ISOTROPY sofware.[47] The crystal lattice parameters obtained in the $P4/nmm$ space group are summarised in Table II; the fitting quality is illustrated by Fig. 12, where the inset present polyhedral representation of the crystal structure. These parameters were then used to refine the magnetic structure from the data recorded at 200 K (E2 diffractometer). The refinement was successful with a correct description of the magnetic intensities (Fig. 13a); the reliability factors are $R_{nuc} = 4.85$ % and $R_{mag} = 9.85$ %. Although there are six nominally independent Co positions in the crystal lattice, the magnetic structure shown schematically in Fig. 13b was refined assuming equal moments in all the pyramidal positions. In this approximation, the ferrimagnetic component results purely from the two non-equivalent sixfold coordinated Co sites. The obtained values of the magnetic moments lying in the (*ab*) plane for the octahedral and pyramidal positions are 1.4(1) $\mu_B$, 0.0(1) $\mu_B$ and 2.0(1) $\mu_B$, respectively. This leads again to the conclusion that the assumed electronic ordering in the octahedral position is the spin-state ordering between diamagnetic and paramagnetic cobalt ions. A comparison of the obtained results with those reported in Ref. 44 for the 332 structural type shows that the predicted type of electronic ordering and the spin configuration are both the same. However, the symmetry was determined incorrectly (*P4mm* instead of *P4/nmm*) due to the structure complexity. In this work, the appropriate space group was deduced from the first principles symmetry approach used for the calculations the ISOTROPY software, thus excluding mistakes. The content of magnetic representations on the cobalt positions in the *P4/nmm* space group was recalculated for $k = 0$

$$D_m^{k=0}(8i) = D_m^{k=0}(8j) = \Gamma_1^+ + \Gamma_2^+ + \Gamma_3^+ + \Gamma_4^+ + 2\Gamma_5^+ + \Gamma_1^- + \Gamma_2^- + \Gamma_3^- + \Gamma_4^- + 2\Gamma_5^-$$

$$D_m^{k=0}(2c) = \Gamma_3^+ + \Gamma_5^+ + \Gamma_1^- + \Gamma_5^- \tag{4}$$

The magnetic structure determined from the neutron diffraction experiment, can be constructed as a linear combination of the basis functions of the single $\Gamma_5^+$ irreducible representation for the *P4/nmm* space group (Table III). This result demonstrates that the proposed magnetic structure is entirely consistent with the symmetry constrains and, hence, is correct with a high probability.

The magnetic measurements (Fig. 3) and the neutron diffraction data (Fig. 10) both show a change of the magnetic structure below 160 K. The new antiferromagnetic reflections can be indexed by doubling the $3\sqrt{2}a_p \times 3\sqrt{2}a_p \times 2a_p$ unit cell along the *c* direction (wave vector $k = c^*/2$). However, some reflections of the high-temperature magnetic phase, in particular (3/2,3/2,1), do not vanish on cooling. This observation contradicts the assumption[44] that the transition into low-temperature magnetic structure occurs via antitranslation along the *c* axis, based on the fact that the wave vector $k = c^*/2$ forbids the appearance of the magnetic reflections with $l = 2n+1$ (in the $3\sqrt{2}a_p \times 3\sqrt{2}a_p \times 4a_p$ unit cell indexation). To remove the discrepancy, a structural phase transition resulting in $3\sqrt{2}a_p \times 3\sqrt{2}a_p \times 4a_p$ unit cell for the low-temperature crystal lattice should accompany the spin structure transformation. In this case, the wave vector for the magnetic structure is again $k = 0$, and the reflections with $l = 2n+1$ are allowed. An examination of the neutron diffraction patterns did not reveal any direct evidences of the structural transformation, suggesting that the symmetry changes may result from a rearrangement in the electronic sublattice. To identify the new type of spin-state ordering and the appropriate symmetry, all isotropy subgroups of the *P4/mmm* space group associated with the A ($k = a^*/2 + b^*/2 + c^*/2$) point were calculated.

Considering only maximal subgroups splitting the 8r cobalt position in the parent structure, four one-dimensional irreducible representations were selected, namely $A_1^+$, $A_4^+$, $A_2^-$ and $A_3^-$; these mediate the transition *P4/mmm* → *I4/mmm* in all cases. In spite of the same resultant symmetry, the crystal structures produced by the distinct irreducible representations are different. The distribution of Co positions in the $3\sqrt{2}a_p \times 3\sqrt{2}a_p \times 4a_p$ unit cell is schematically shown in Figure 14, where only two variants are presented since the location of non-equivalent cobalt sites is the same for the $A_2^- - A_4^+$ and $A_1^+ - A_3^-$ pairs. These structures, however, are different in respect of the Ba/Y independent positions.

By further application of isotropic negative exchange interactions, the two magnetic structures were obtained and preliminary tested to compare the calculated and experimental magnetic intensity values. The model conjugated with the crystal structure shown in Fig. 14a gives a good agreement. In this model, the spin-state ordering in the [CoO$_2$] planes is similar to that found in the high temperature *P4/nmm* phase. The difference is in the alternation of the spin-state ordered planes along the *c* direction. In case of the *P4/nmm* symmetry, the neighbouring [CoO$_2$] planes are shifted along the planar diagonal in a half-period ($\cdots$ [CoO$_2$]$'$ $\cdots$ [CoO$_2$]$''$ $\cdots$ [CoO$_2$]$'$ $\cdots$ sequence), whereas for *I4/mmm* this shift appears once per two planes, which results in a quadrupling of the *c* parameter ($\cdots$ [CoO$_2$]$'$ $\cdots$ [CoO$_2$]$'$ $\cdots$ [CoO$_2$]$''$ $\cdots$ [CoO$_2$]$''$ $\cdots$ sequence). The $A_2^-$ and $A_4^+$ representations produce the structures where the shifted [CoO$_2$]$'$ / [CoO$_2$]$''$ planes are divided by [YO$_{0.44}$] and [BaO] layers, respectively. These two cases cannot be unambiguously distinguished by refining either crystal or magnetic structure. The model associated with the $A_2^-$ irreducible representation was chosen for definiteness. The structural parameters at T = 100 K, refined using this model, are listed in Table IV; fitting quality is illustrated in Figure 15, presenting also the polyhedral representation of the lattice.

The refinement of the spin structure was constrained to vary only three independent magnetic moments, two for octahedral and one for pyramidal cobalt. The obtained values at T = 100 K are 2.3(1) $\mu_B$, 0.1(1) $\mu_B$ and 2.6(1) $\mu_B$, respectively. The refinement quality is satisfactory (Fig. 16a) with the reliability factors $R_{nuc}$ = 5.41%, $R_{mag}$ = 11.6%. The directions of the magnetic moments in the (*ab*) planes of $3\sqrt{2}a_p \times 3\sqrt{2}a_p \times 4a_p$ unit cell are shown in Figure 16b. The magnetic moments are not saturated yet, but their values are too large already for the intermediate-spin state ($t_{2g}^5 e_g^1$ S = 1). In combination with isotropic negative superexchange interactions between paramagnetic Co ions, this result is consistent with the high-spin electronic configuration.

The obtained antiferromagnetic spin structure suggests a different mechanism of the phase transition at $T_i$ in comparison with that discussed above for YBaCo$_2$O$_{5.5}$. The most reasonable explanation relates to a competition between the elastic energy and exchange interactions. The former requires all neighbouring cobalt ions in octahedral position to have different spin states, when a gain in the elastic energy is maximum. Such situation takes place in the high-temperature *P4/nmm* phase. However, in this structure, octahedrally coordinated Co ions having nonzero magnetic moment are only exchange-coupled with two paramagnetic Co ions (in pyramidal positions). With decreasing temperature, the role of exchange interactions increases, and the transition into *I4/mmm* phase becomes favourable since, in this phase, each paramagnetic sixfold coordinated cobalt ions is exchange-coupled with three other magnetoactive Co ions (two pyramidally and one octahedrally coordinated). Therefore, this type of the spin-state ordering is preferable in respect of the exchange energy. A similar situation is likely in $YBaCo_2O_{5.5}$ at low temperatures, below $T_L \sim 190$ K.

Finally, the basis functions for irreducible representations of the wave vector group were calculated, and it was found that the proposed magnetic structure is not consistent with the *I4/mmm* symmetry. None linear combination of the basis functions of a single irreducible representation may produce the obtained magnetic ordering, either pyramidal or octahedral sublattices. This situation is not, however, surprising in the present case, and some symmetry motives can be involved to resolve the problem. The *I4/mmm* ($3\sqrt{2}a_p \times 3\sqrt{2}a_p \times 4a_p$) symmetry is caused by the spin-state ordering in the 8*r* octahedral position of the parent *P4/mmm* ($3a_p \times 3a_p \times 2a_p$) structure, whilst the electronic configuration of pyramidally coordinated Co ions remains fixed at any temperature. One can suppose that the symmetry of the magnetic structure is determined by the pyramidal sublattice symmetry, which remains the same as in the parent *P4/mmm* structure. The splitting of the 8*t* and 2*c* pyramidal positions due to the phase transformation is effective; no electronic rearrangement in these sites is involved. Therefore, the spin ordering in these positions meet the requirements of the *P4/mmm* symmetry. The dominant pyramidal sublattice with higher symmetry tunes the spin ordering in the octahedral sites, resulting in breakdown of the *I4/mmm* symmetry constrains.

**Conclusions**

The crystal structure of $YBaCo_2O_{5.50}$ has an orthorhombic *Pmma* symmetry with $2a_p \times 2a_p \times 2a_p$ unit cell in the temperature range 190 < T < 295 K. Two non-equivalent 2*e* octahedral positions are occupied by diamagnetic (low-spin) and paramagnetic (high-spin state) $Co^{3+}$ ions, forming chess-board like spin-state ordered (*ac*) planes. The fivefold coordinated Co ions in two independent 2*f* positions adopt the fixed high-spin electronic configuration. A long-range magnetic ordering appears below $T_C \sim 290$ K. The magnetic structure above $T_i \sim 260$ K has the wave vector $k = 0$ and involves isotropic negative exchange interactions between the paramagnetic Co ions in both octahedral and pyramidal coordination. The spontaneous magnetisation originates from the spin-state ordering in the octahedral sublattice. At $T_i$, a magnetic phase transition to antiferromagnetic structure with the wave vector $k = c^*/2$ occurs. In the low temperature magnetic phase, the coupling between the (*ab*) planes divided by [BaO] and [YO$_{0.52}$] layers is negative and positive, respectively. The driving force for the magnetic phase transition is a thermal excitation of magnetic moment on the diamagnetic cobalt ions. In the low-temperature spin configuration, these excitations result in inevitable magnetic frustrations which may be avoided by transition to the structure with $k = 0$. Below $T_L \sim 190$ K, a new type of antiferromagnetic ordering is formed. The rearrangement of the spin structure is accompanied with the change of the crystal structure symmetry ($2a_p \times 2a_p \times 2a_p \rightarrow 2a_p \times 2a_p \times 4a_p$), apparently due to development of a new type of the spin-state ordered superlattice.

YBaCo$_2$O$_{5.44}$ exhibits a translationally symmetric distribution of extra oxygen anions in the [YO$_{0.44}$] layers, resulting in tetragonal *P4/mmm* symmetry with $3a_p \times 3a_p \times 2a_p$ unit cell dimension. The spin-state ordering between diamagnetic and paramagnetic cobalt ions in the octahedral 8*r* position changes the symmetry to *P4/nmm* ($3\sqrt{2}a_p \times 3\sqrt{2}a_p \times 2a_p$). In the range 160 K<T<290 K, magnetic structure with the wave vector $k = 0$ is formed. In this spin-ordered configuration, all neighbour cobalt ions with nonzero magnetic moments are coupled antiferromagnetically. The spontaneous magnetisation originates from two non-equivalent 8*i* octahedral positions, occupied by cobalt ions in different spin states. The fivefold coordinated ions in four nominally independent Co positions adopt uniform high-spin electronic configuration. The phase transition at $T_i \sim 160$ K involves a reorganisation in both spin-state ordering and magnetic structure. The crystal lattice symmetry of the low-temperature phase is *I4/mmm* ($3\sqrt{2}a_p \times 3\sqrt{2}a_p \times 4a_p$). The wave vector of the conjugated magnetic structure is $k = 0$. The spin structure reorganisation takes place only in the octahedral sublattice, whereas the magnetic ordering in the pyramidal subsystem does not change with the phase transition. The driving force of this transformation is a competition between the elastic energy and exchange interactions. In the high-temperature magnetic phase, sixfold coordinated Co ions with nonzero magnetic moment are surrounded by four diamagnetic and two paramagnetic neighbours, which is favourable in respect of the elastic energy. In the low-temperature magnetic phase, these ions have three diamagnetic and three paramagnetic neighbours, resulting in exchange energy gain. A similar mechanism may be responsible for the phase transformation in YBaCo$_2$O$_{5.5}$ at $T_L \sim 190$ K.

**Captions of the figures**

**Fig. 1** Ordering of extra oxygen ions in [LnO$_\delta$] layers (○ Ln, ● O); a) $3a_p \times 3a_p \times 2a_p$ (332) type of superstructure at $\delta=0.44$, b) $a_p \times 2a_p \times 2a_p$ (122) type of superstructure at $\delta=0.50$ and c) $2a_p \times 2a_p \times 2a_p$ (222) type of superstructure at $\delta=0.75$.

**Fig. 2** Variations of oxygen nonstoichiometry in YBaCo$_2$O$_{5+\delta}$ samples on heating in air. Inset illustrates relative weight change of YBaCo$_2$O$_{5+\delta}$ sample upon reduction at 1223-1323 K.

**Fig. 3** Zero field cooling magnetization as a function of temperature for the YBaCo$_2$O$_{5.5}$ and YBaCo$_2$O$_{5.44}$ compositions, registered in H = 10 kOe.

**Fig. 4** E2 neutron diffraction patterns obtained at different temperatures from the YBaCo$_2$O$_{5.5}$ composition. The indexation is given in $a_p \times 2a_p \times 2a_p$ unit cell.

**Fig. 5** Refined in *Pmma* space group E9 neutron diffraction pattern ($\lambda = 1.7974$ Å) from the YBaCo$_2$O$_{5.5}$ composition, taken at 293 K. Inset shows a polyhedral representation of the corresponding crystal structure. Two non-equivalent octahedral Co positions are presented by the light and dark grey polyhedra, respectively.

**Fig. 6** Refined E2 diffraction pattern ($\lambda = 2.39$ Å) from the YBaCo$_2$O$_{5.5}$ composition, taken at 275 K. Inset demonstrates expanded view of the $10 - 45°\ 2\Theta$ range.

**Fig. 7** Schematic representation of the magnetic structure above T$_i$ (a) and below T$_i$ (b) for YBaCo$_2$O$_{5.5}$ (Co1, Co2 and Co3, Co4 - pyramidal and octahedral positions, respectively). Diamagnetic Co ions in the low-spin state are denoted by gray circles. The (c) and (d) panels demonstrate the corresponding (*ac*) plains containing sixfold coordinated cobalt. Exited (virtual) magnetic moments on the low-spin Co ions are shown by the dot arrows.

**Fig. 8** Refined E2 diffraction pattern ($\lambda = 2.39$ Å) from the YBaCo$_2$O$_{5.5}$ composition, taken at 250 K. Inset demonstrates expanded view of the $5 - 45°\ 2\Theta$ range.

**Fig. 9** Schematic representation of the 332 structural type (*P4/mmm*, $3a_p \times 3a_p \times 2a_p$) (a). Displacements of yttrium (●) and oxygen (○) ions from their highly symmetric positions in [YO$_{0.44}$] layers, corresponding to two different minima (b and c) in the refinement of the crystal structure of YBaCo$_2$O$_{5.44}$.

**Fig. 10** E2 neutron diffraction patterns obtained at different temperatures from the YBaCo$_2$O$_{5.44}$ composition. The indexation is given in $3a_p \times 3a_p \times 2a_p$ unit cell.

**Fig. 11** (*ab*) planes in $3\sqrt{2}a_p \times 3\sqrt{2}a_p \times 2a_p$ unit cell of the different isotropy subgroups, containing Co ions in octahedral (large circles) and pyramidal (small grey circles) coordination. Two non-equivalent octahedral positions are presented by black and white circles, respectively. Positive

and negative directions of magnetic moments on the sixfold coordinated Co ions in the *G* type mode are also presented.

**Fig. 12** Refined in the *P4/nmm* space group E9 neutron diffraction pattern ($\lambda = 1.7974$ Å) from the YBaCo$_2$O$_{5.44}$ composition, taken at 200 K. Inset shows a polyhedral representation of the corresponding crystal structure (*P4/nmm* origin choice 1). The two non-equivalent octahedral Co positions are presented by the light and dark grey polyhedra, respectively.

**Fig. 13** Refined E2 diffraction pattern ($\lambda = 2.39$ Å) from the YBaCo$_2$O$_{5.44}$ composition, taken at 200 K (a). Inset demonstrates expanded view of the $5 - 20^\circ$ $2\Theta$ range. Positive and negative directions of the magnetic moments on paramagnetic Co ions in the neighbour (*ab*) planes in $3\sqrt{2}a_p \times 3\sqrt{2}a_p \times 2a_p$ unit cell (b). The diamagnetic sixfold coordinated Co ions are shown by the black circles.

**Fig. 14** (*ab*) planes in $3\sqrt{2}a_p \times 3\sqrt{2}a_p \times 4a_p$ unit cell of the different isotropy subgroups, containing Co ions in octahedral (large circles) and pyramidal (small grey circles) coordination. Two non-equivalent octahedral positions are presented by black and white circles, respectively.

**Fig. 15** Refined in *I4/mmm* space group E9 neutron diffraction pattern ($\lambda = 1.7974$ Å) from the YBaCo$_2$O$_{5.44}$ composition, taken at 100 K. Inset shows a polyhedral representation of the corresponding crystal structure. The non-equivalent octahedral Co positions are presented by the light and dark grey polyhedra, respectively.

**Fig. 16** Refined E2 diffraction pattern ($\lambda = 2.39$ Å) from the YBaCo$_2$O$_{5.44}$ composition, taken at 100 K (a). Inset demonstrates expanded view of the $5 - 20^\circ$ $2\Theta$ range. Positive and negative directions of the magnetic moments on paramagnetic Co ions in the (*ab*) planes of $3\sqrt{2}a_p \times 3\sqrt{2}a_p \times 4a_p$ unit cell (b). The diamagnetic sixfold coordinated Co ions are shown by the black circles.

**TABLE I.** Atomic positions $x$, $y$, $z$, isotropic temperature factors $B$, and position occupancies $n$, for $YBaCo_2O_{5.5}$ refined in *Pmma* space group from high resolution powder neutron diffraction. The unit cell parameters and reliability factors are: T = 293 K, $a$ = 7.6967(2) Å, $b$ = 7.8141(2) Å, $c$ = 7.5052(2) Å, $R_p$ = 5.30 %, $R_{wp}$ = 6.86 %, $\chi^2$ = 2.23; T = 275 K, $a$ = 7.7012(2) Å, $b$ = 7.8085(2) Å, $c$ = 7.5025(2) Å, $R_p$ = 5.38 %, $R_{wp}$ = 6.96 %, $\chi^2$ = 2.29; T = 250 K, $a$ = 7.7040(2) Å, $b$ = 7.8031(2) Å, $c$ = 7.5004(2) Å, $R_p$ = 5.68 %, $R_{wp}$ = 7.31 %, $\chi^2$ = 2.55.

| T | 293 K | | | | | 275 K | | | | | 250 K | | | | |
| --- | --- | --- | --- | --- | --- | --- | --- | --- | --- | --- | --- | --- | --- | --- | --- |
| Atom | Wyck | $x$ | $y$ | $z$ | $B$ | $n$ | $x$ | $y$ | $z$ | $B$ | $n$ | $x$ | $y$ | $z$ | $B$ | $n$ |
| Y | 4h | 0 | 0.2695(6) | 0.5 | 1.01(7) | 1 | 0 | 0.2696(7) | 0.5 | 1.10(8) | 1 | 0 | 0.2708(6) | 0.5 | 1.06(8) | 1 |
| Ba | 4g | 0 | 0.2523(9) | 0 | 0.8(1) | 1 | 0 | 0.2525(9) | 0 | 0.7(1) | 1 | 0 | 0.251(1) | 0 | 0.5(1) | 1 |
| Co$_{Py1}$ | 2e | 0.25 | 0 | 0.260(2) | 0.4(1) | 1 | 0.25 | 0 | 0.259(2) | 0.5(1) | 1 | 0.25 | 0 | 0.257(2) | 0.5(1) | 1 |
| Co$_{Py2}$ | 2e | 0.25 | 0 | -0.260(2) | 0.4(1) | 1 | 0.25 | 0 | -0.259(2) | 0.5(1) | 1 | 0.25 | 0 | -0.257(2) | 0.5(1) | 1 |
| Co$_{Oc1}$ | 2f | 0.25 | 0.5 | 0.250(2) | 0.4(1) | 1 | 0.25 | 0.5 | 0.250(2) | 0.5(1) | 1 | 0.25 | 0.5 | 0.252(2) | 0.5(1) | 1 |
| Co$_{Oc2}$ | 2f | 0.25 | 0.5 | -0.250(2) | 0.4(1) | 1 | 0.25 | 0.5 | -0.250(2) | 0.5(1) | 1 | 0.25 | 0.5 | -0.251(2) | 0.5(1) | 1 |
| O1 | 2e | 0.25 | 0 | -0.001(3) | 0.6(2) | 1 | 0.25 | 0 | -0.002(3) | 0.6(2) | 1 | 0.25 | 0 | -0.004(3) | 0.7(2) | 1 |
| O2 | 4i | 0.017(1) | 0 | 0.3131(8) | 0.5(1) | 1 | 0.019(1) | 0 | 0.3128(8) | 0.9(1) | 1 | 0.022(1) | 0 | 0.3138(8) | 0.7(1) | 1 |
| O3 | 4k | 0.25 | -0.234(1) | -0.302(2) | 0.9(2) | 1 | 0.25 | -0.230(1) | -0.303(2) | 0.8(2) | 1 | 0.25 | -0.231(1) | -0.303(2) | 0.7(2) | 1 |
| O4 | 4k | 0.25 | 0.257(1) | 0.295(2) | 1.6(3) | 1 | 0.25 | 0.258(1) | 0.292(2) | 1.5(2) | 1 | 0.25 | 0.260(1) | 0.292(2) | 1.3(2) | 1 |
| O5 | 2f | 0.25 | 0.5 | -0.000(4) | 1.3(2) | 1 | 0.25 | 0.5 | 0.001(3) | 1.4(2) | 1 | 0.25 | 0.5 | 0.004(3) | 1.2(2) | 1 |
| O6 | 4j | -0.003(2) | 0.5 | 0.271(1) | 2.0(1) | 1 | -0.004(2) | 0.5 | 0.272(1) | 2.0(1) | 1 | -0.002(2) | 0.5 | 0.270(1) | 2.0(1) | 1 |
| O7 | 2f | 0.25 | 0.5 | 0.494(4) | 1.2(3) | 0.96(2) | 0.25 | 0.5 | 0.499(4) | 1.0(3) | 0.92(2) | 0.25 | 0.5 | 0.496(3) | 1.1(3) | 0.92(2) |
| O71 | 2e | 0.25 | 0 | 0.5 | 0.6(2) | 0.00(1) | 0.25 | 0 | 0.5 | 0.6(2) | 0.00(1) | 0.25 | 0 | 0.5 | 0.7(1) | 0.00(1) |

**TABLE II.** Atomic positions $x$, $y$, $z$, isotropic temperature factors $B$, and position occupancies, for $YBaCo_2O_{5.44}$ refined in the $P4/nmm$ space group (origin choice 2) from the high resolution powder neutron diffraction. The unit cell parameters and reliability factors are: T = 295 K, $a$ = 16.4285(2) Å, $b$ = 16.4285(2) Å, $c$ = 7.4967(1) Å, $R_p$ = 5.79 %, $R_{wp}$ = 7.36 %, $\chi^2$ = 1.90; T = 200 K, $a$ = 16.4143(2) Å, $b$ = 16.4143(2) Å, $c$ = 7.4855(1) Å, $R_p$ = 5.13 %, $R_{wp}$ = 6.61 %, $\chi^2$ = 2.9.

| T | | 295 K | | | | | 200 K | | | | |
|---|---|---|---|---|---|---|---|---|---|---|---|
| Atom | Wyck | $x$ | $y$ | $z$ | $B$ | Occup. | $x$ | $y$ | $z$ | $B$ | Occup. |
| Ba1 | 2a | 0.75 | 0.25 | 0 | 0.4(1) | 1 | 0.75 | 0.25 | 0 | 0.2(1) | 1 |
| Ba2 | 8i | 0.25 | -0.587(1) | -0.016(3) | 0.4(1) | 1 | 0.25 | -0.587(1) | -0.019(2) | 0.2(1) | 1 |
| Ba3 | 8g | 0.5866(7) | -0.5866(7) | 0 | 0.4(1) | 1 | 0.5872(6) | -0.5872(6) | 0 | 0.2(1) | 1 |
| Y1 | 2b | 0.75 | 0.25 | 0.5 | 1.1(1) | 1 | 0.75 | 0.25 | 0.5 | 0.96(8) | 1 |
| Y2 | 8h | 0.5948(4) | -0.5948(4) | 0.5 | 1.1(1) | 1 | 0.5951(4) | -0.5951(4) | 0.5 | 0.96(8) | 1 |
| Y3 | 8i | 0.25 | -0.5767(8) | 0.498(3) | 1.1(1) | 1 | 0.25 | -0.5767(7) | 0.497(2) | 0.96(8) | 1 |
| Co$_{Oc1}$ | 8i | 0.25 | -0.078(3) | 0.250(3) | 0.3(2) | 1 | 0.25 | -0.073(2) | 0.253(2) | 0.3(1) | 1 |
| Co$_{Oc2}$ | 8i | 0.25 | -0.403(3) | -0.250(3) | 0.3(2) | 1 | 0.25 | -0.408(2) | -0.253(2) | 0.3(1) | 1 |
| Co$_{Py1}$ | 8j | -0.082(1) | -0.082(1) | -0.260(2) | 0.3(2) | 1 | -0.0812(9) | -0.0812(9) | -0.257(2) | 0.3(1) | 1 |
| Co$_{Py2}$ | 8j | -0.082(1) | -0.082(1) | 0.260(2) | 0.3(2) | 1 | -0.0812(9) | -0.0812(9) | -0.257(2) | 0.3(1) | 1 |
| Co$_{Py3}$ | 2c | 0.25 | 0.25 | 0.260(2) | 0.3(2) | 1 | 0.25 | 0.25 | 0.257(2) | 0.3(1) | 1 |
| Co$_{Py4}$ | 2c | 0.25 | 0.25 | -0.260(2) | 0.3(2) | 1 | 0.25 | 0.25 | -0.257(2) | 0.3(1) | 1 |
| O1 | 8i | 0.25 | -0.427(1) | 0.007(3) | 0.6(2) | 1 | 0.25 | -0.4273(9) | 0.006(3) | 0.6(2) | 1 |
| O1 | 8i | 0.25 | -0.427(1) | 0.007(3) | 0.6(2) | 1 | 0.25 | -0.4273(9) | 0.006(3) | 0.6(2) | 1 |
| O2 | 8j | -0.0810(7) | -0.0810(7) | 0.002(4) | 0.7(3) | 1 | -0.0814(6) | -0.0814(6) | 0.006(3) | 0.4(2) | 1 |
| O3 | 2c | 0.25 | 0.25 | 0.012(9) | 1.7(8) | 1 | 0.25 | 0.25 | 0.002(9) | 1.9(7) | 1 |
| O4 | 16k | -0.0007(9) | -0.665(1) | 0.305(2) | 0.8(3) | 1 | 0.0046(8) | -0.6633(9) | 0.304(2) | 0.4(2) | 1 |
| O5 | 16k | -0.5043(9) | -0.666(1) | -0.304(2) | 0.5(3) | 1 | -0.5013(9) | -0.668(1) | -0.304(2) | 0.5(2) | 1 |
| O6 | 16k | -0.167(1) | -0.839(1) | 0.262(1) | 1.6(2) | 1 | -0.1701(9) | -0.8394(9) | 0.2598(9) | 1.5(2) | 1 |
| O7 | 8j | 0.004(2) | 0.004(2) | -0.310(3) | 1.8(4) | 1 | 0.001(1) | 0.001(1) | -0.311(2) | 1.6(3) | 1 |
| O8 | 8j | -0.172(1) | -0.172(1) | -0.299(3) | 1.9(5) | 1 | -0.167(1) | -0.167(1) | -0.301(3) | 1.9(5) | 1 |
| O9 | 8j | -0.1647(9) | -0.1647(9) | 0.326(2) | 0.4(3) | 1 | -0.1660(8) | -0.1660(8) | 0.324(2) | 0.2(3) | 1 |
| O10 | 8i | 0.25 | -0.3989(7) | 0.490(4) | 0.5(3) | 0.92(2) | 0.25 | -0.3988(8) | 0.488(3) | 0.5(3) | 0.91(2) |
| O11 | 8j | -0.25(-) | -0.25(-) | 0.5 | 0.5(3) | 0.04(2) | -0.25(-) | -0.25(-) | 0.5 | 0.5(3) | 0.04(2) |

**TABLE III.** Basis functions, $S^k$, of the $\Gamma_5^+$ irreducible representation of the $P4/nmm$ space group for the octahedral and pyramidal sublattices.

| Wyckoff | | 2c | | 8i | | | | | | | | 8j | | | | | | | |
|---|---|---|---|---|---|---|---|---|---|---|---|---|---|---|---|---|---|---|---|
| $i^a$ | | 1 | 2 | 1 | 2 | 3 | 4 | 5 | 6 | 7 | 8 | 1 | 2 | 3 | 4 | 5 | 6 | 7 | 8 |
| $n^b = 1$ | $S_x^k$ | $\bar{1}$ 0 | $\bar{1}$ 0 | 1 0 | 1 0 | 1 0 | 1 0 | 0 0 | 0 0 | 0 0 | 0 0 | 1 $\bar{1}$ | 1 $\bar{1}$ | 1 $\bar{1}$ | 1 $\bar{1}$ | $\bar{1}$ 1 | $\bar{1}$ 1 | $\bar{1}$ 1 | $\bar{1}$ 1 |
| | $S_y^k$ | 0 $\bar{1}$ | 0 $\bar{1}$ | 0 0 | 0 0 | 0 0 | 0 0 | 0 1 | 0 1 | 0 1 | 0 1 | $\bar{1}$ 1 | $\bar{1}$ 1 | $\bar{1}$ 1 | $\bar{1}$ 1 | 1 $\bar{1}$ | 1 $\bar{1}$ | 1 $\bar{1}$ | 1 $\bar{1}$ |
| | $S_z^k$ | 0 0 | 0 0 | 0 0 | 0 0 | 0 0 | 0 0 | 0 0 | 0 0 | 0 0 | 0 0 | 0 0 | 0 0 | 0 0 | 0 0 | 0 0 | 0 0 | 0 0 | 0 0 |
| $n = 2$ | $S_x^k$ | | | 0 0 | 0 0 | 0 0 | 0 0 | 0 0 | 1 0 | 1 0 | 1 0 | 1 1 | 1 $\bar{1}$ | 1 $\bar{1}$ | 1 1 | 1 1 | 1 $\bar{1}$ | 1 $\bar{1}$ | 1 1 |
| | $S_y^k$ | | | 0 1 | 0 1 | 0 1 | 0 1 | 0 1 | 0 0 | 0 0 | 0 0 | 1 1 | $\bar{1}$ 1 | $\bar{1}$ 1 | 1 1 | 1 1 | $\bar{1}$ 1 | $\bar{1}$ 1 | 1 1 |
| | $S_z^k$ | | | 0 0 | 0 0 | 0 0 | 0 0 | 0 0 | 0 0 | 0 0 | 0 0 | 0 0 | 0 0 | 0 0 | 0 0 | 0 0 | 0 0 | 0 0 | 0 0 |

[a] $i$ numerates cobalt ions in the 2c, 8i and 8j positions.

[b] $n$ numerates sets of basis functions of the $\Gamma_5^+$ irreducible representation entering two times the magnetic representation on 8i and 8j positions.

**P4/nmm, origin choice 1**. 2c - 1(-1/2,0,z), 2(0,1/2,-z); 8i - 1(-1/2,x-1/4,z), 2(0,-x+3/4,-z), 3(1,x+1/4,-z), 4(1/2,-x+1/4,z), 5(-x+1/4,1/2,-z), 6(-x+3/4,0,z), 7(x+1/4,1,z), 8(x-1/4,-1/2,-z); 8j - 1(x-3/4,x-1/4,z), 2(x-1/4,-x+3/4,-z), 3(-x+5/4,x+1/4,-z), 4(-x+3/4,-x+1/4,z), 5(-x+1/4,-x+3/4,-z), 6(-x+3/4,x-1/4,z), 7(x+1/4,-x+5/4,z), 8(x-1/4,x-3/4,-z).

**P4/nmm, origin choice 2**. 2c - 1(1/4,1/4,z), 2(3/4,3/4,-z); 8i - 1(1/4,x,z), 2(3/4,-x+1,-z), 3(7/4,x+1/2,-z), 4(5/4,-x+1/2,z), 5(-x+1,3/4,-z), 6(-x+3/2,1/4,z), 7(x+1,5/4,z), 8(x+1/2,-1/4,-z); 8j - 1(x,x,z), 2(x+1/2,-x+1,-z), 3(-x+2,x+1/2,-z), 4(-x+3/2,-x+1/2,z), 5(-x+1,-x+1,-z), 6(-x+3/2,x,z), 7(x+1,-x+3/2,z), 8(x+1/2,x-1/2,-z).

**TABLE IV.** Atomic positions $x$, $y$, $z$, isotropic temperature factors $B$, and position occupancies, for YBaCo$_2$O$_{5.44}$ refined in the *I4/mmm* space group from the high resolution powder neutron diffraction data registered at T = 100 K. The unit cell parameters and reliability factors are, $a = b = 16.4004(1)$ Å, $c = 14.9494(2)$ Å, $R_p = 5.17$ %, $R_{wp} = 6.87$ %, $\chi^2 = 3.32$.

| Atom | Position | $x$ | $y$ | $z$ | B | Occup. |
|---|---|---|---|---|---|---|
| Ba1 | 4d | 0 | 0.5 | 0.25 | 0.1(1) | 1 |
| Ba2 | 16n | 0 | 0.163(1) | 0.256(2) | 0.1(1) | 1 |
| Ba3 | 16k | -0.3365(6) | 0.1635(6) | 0.25 | 0.1(1) | 1 |
| Y1 | 4c | 0 | 0.5 | 0 | 0.8(1) | 1 |
| Y2 | 16l | 0.842(1) | -0.346(1) | 0 | 0.8(1) | 1 |
| Y3 | 8i | 0.831(2) | 0 | 0 | 0.8(1) | 1 |
| Y4 | 8j | 0.322(2) | 0.5 | 0 | 0.8(1) | 1 |
| Co$_{Oc1}$ | 16n | 0 | 0.322(2) | 0.121(1) | 0.2(1) | 1 |
| Co$_{Oc2}$ | 16n | 0 | 0.660(3) | 0.379(1) | 0.2(1) | 1 |
| Co$_{Py1}$ | 4e | 0 | 0 | 0.8755(9) | 0.2(1) | 1 |
| Co$_{Py2}$ | 4e | 0 | 0 | 0.3755(9) | 0.2(1) | 1 |
| Co$_{Py3}$ | 16m | 0.166(1) | 0.166(1) | 0.12445(9) | 0.2(1) | 1 |
| Co$_{Py4}$ | 16m | 0.166(1) | 0.166(1) | -0.3755(9) | 0.2(1) | 1 |
| O1 | 16n | 0 | 0.323(1) | 0.248(3) | 0.8(2) | 1 |
| O2 | 16m | 0.1686(6) | 0.1686(6) | -0.252(2) | 0.2(2) | 1 |
| O3 | 4e | 0 | 0 | 0.75 | 2.2(8) | 1 |
| O4 | 32o | 0.2547(8) | 0.5808(8) | 0.405(1) | -0.1(3) | 1 |
| O5 | 32o | -0.2528(9) | 0.5894(8) | 0.098(1) | 0.4(3) | 1 |
| O6 | 32o | 0.085(1) | 0.414(1) | 0.3790(5) | 1.2(2) | 1 |
| O7 | 16m | 0.251(1) | 0.251(1) | -0.404(1) | 1.2(3) | 1 |
| O8 | 16m | 0.082(1) | 0.082(1) | 0.097(2) | 2.0(7) | 1 |
| O9 | 16m | 0.084(1) | 0.084(1) | -0.409(1) | 0.0(4) | 1 |
| O10 | 8i | 0.647(2) | 0 | 0 | -0.1(3) | 0.88(4) |
| O11 | 8j | 0.153(3) | 0.5 | 0 | -0.1(3) | 0.88(4) |
| O12 | 8h | 0.1666 | 0.1666 | 0 | -0.1(3) | 0.08(4) |
| O13 | 8h | 0.3333 | 0.3333 | 0 | -0.1(3) | 0.08(4) |

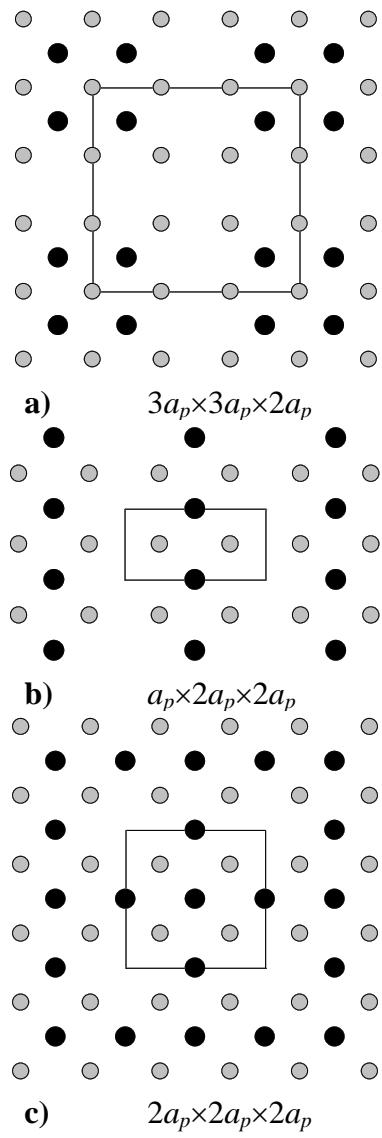

**a)**   $3a_p \times 3a_p \times 2a_p$

**b)**   $a_p \times 2a_p \times 2a_p$

**c)**   $2a_p \times 2a_p \times 2a_p$

**Fig. 1**

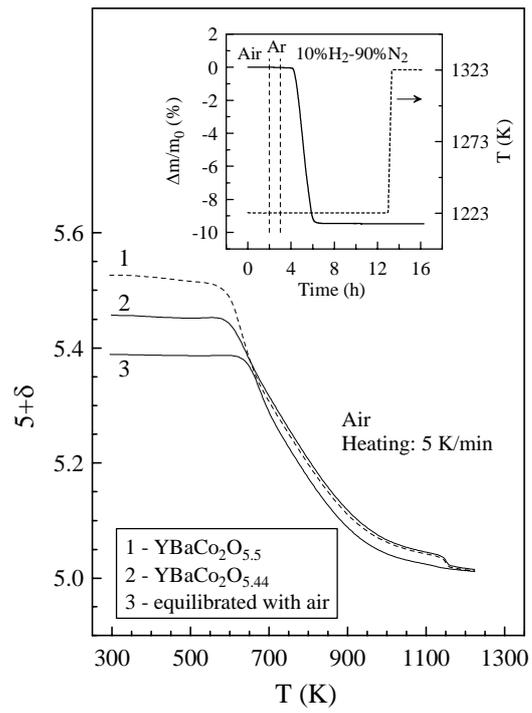

**Fig. 2**

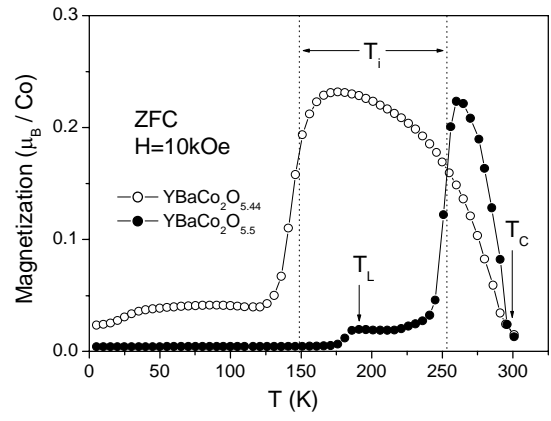

**Fig. 3**

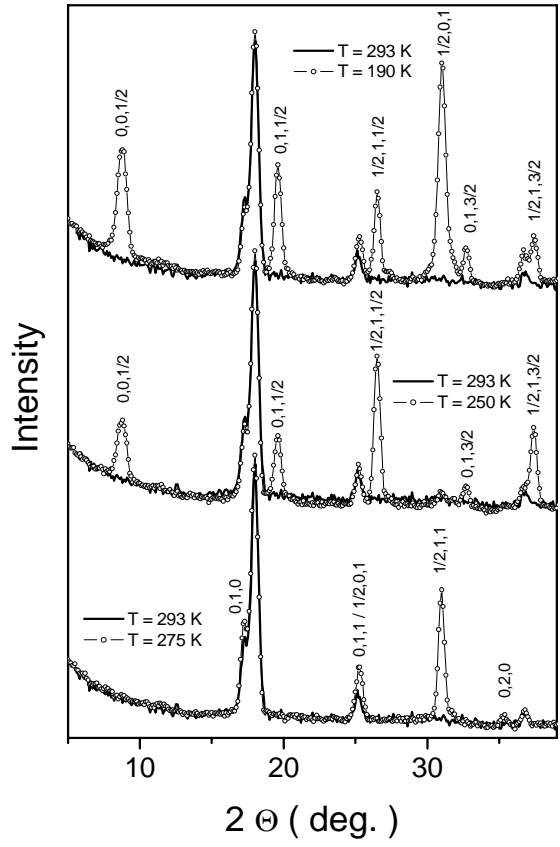

**Fig. 4**

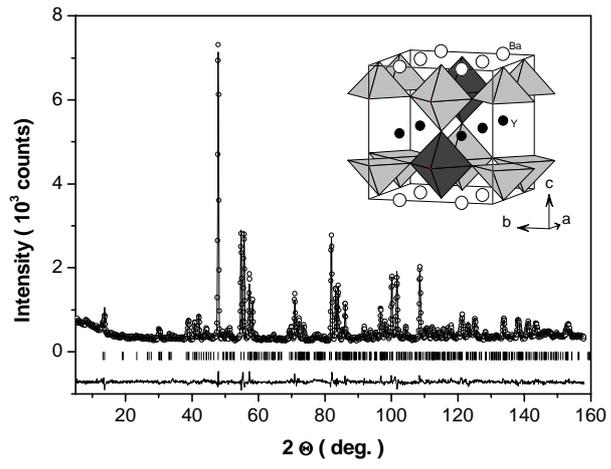

**Fig. 5**

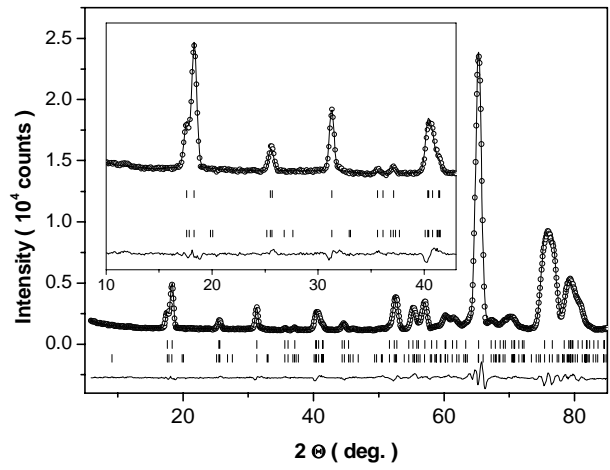

**Fig. 6**

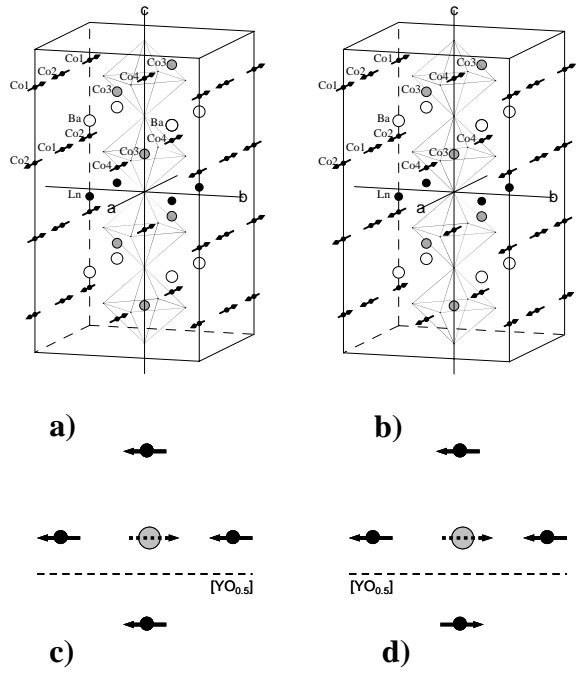

**Fig. 7**

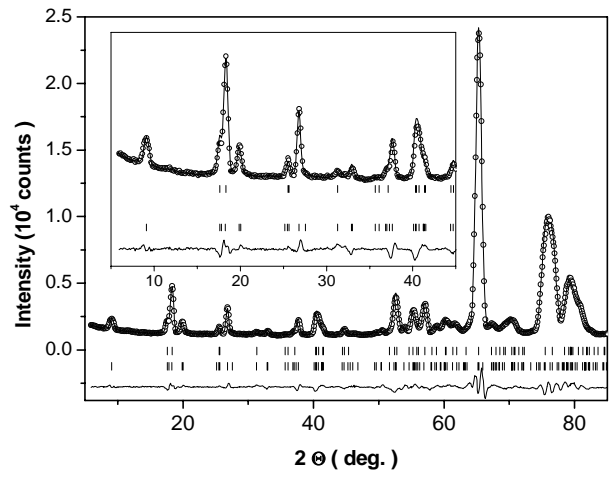

**Fig. 8**

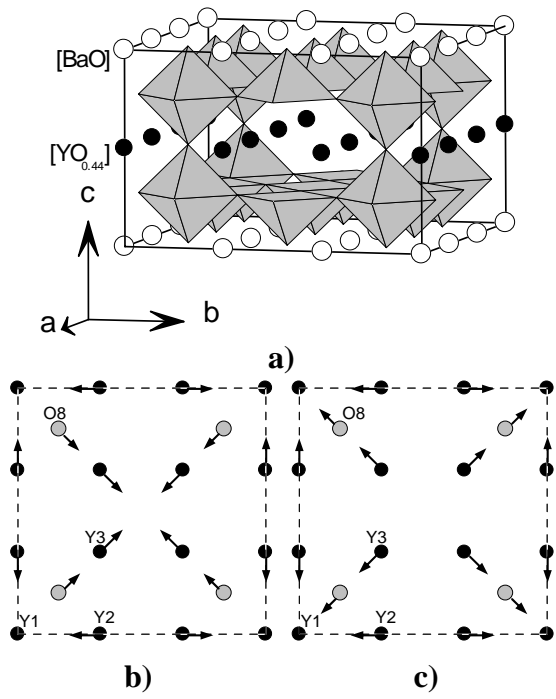

**Fig. 9**

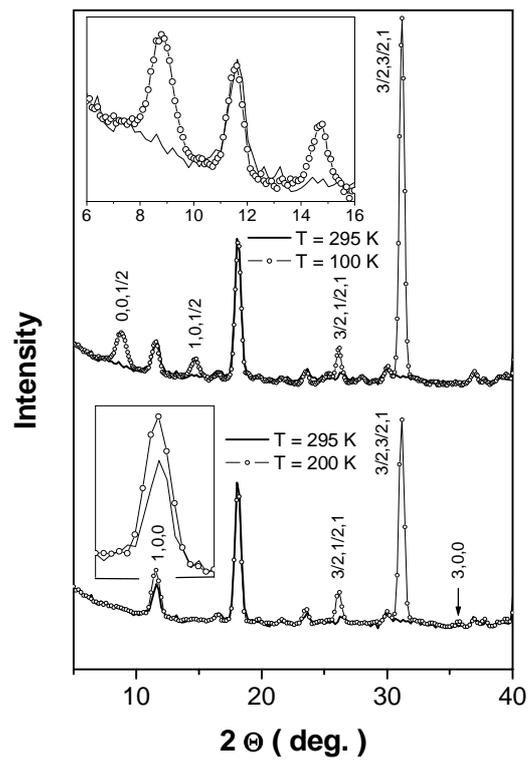

**Fig. 10**

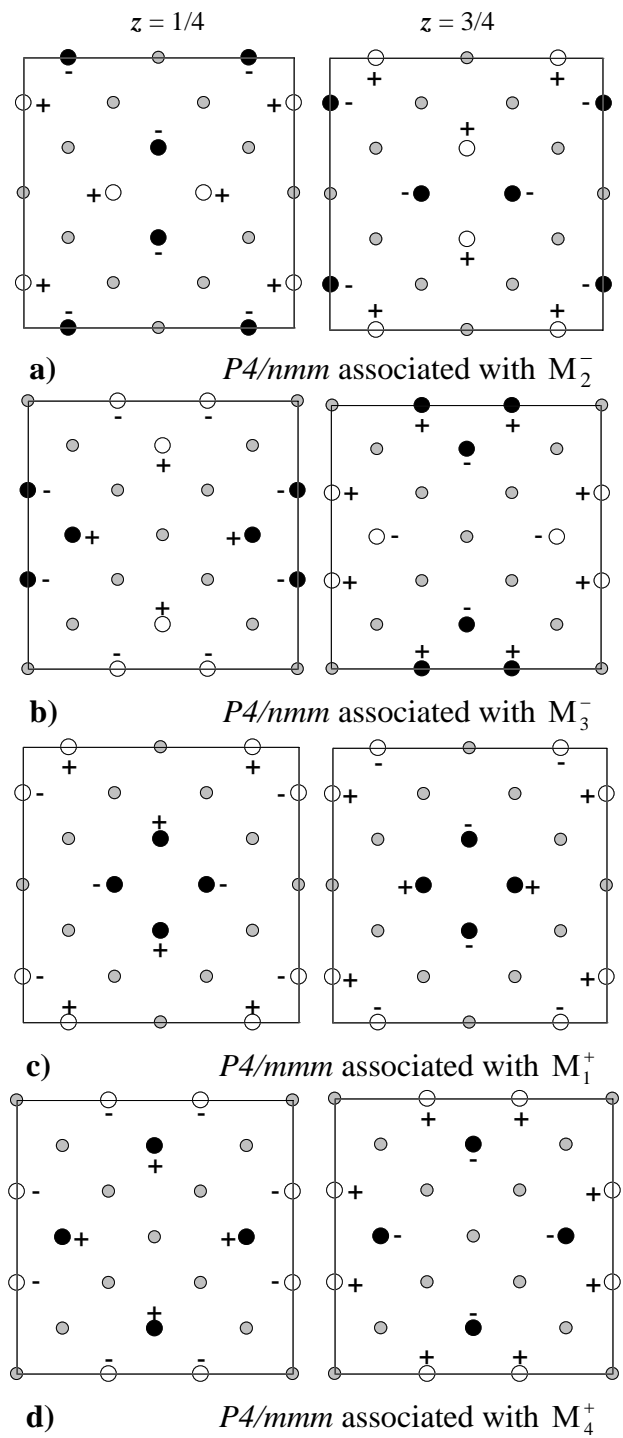

**Fig. 11**

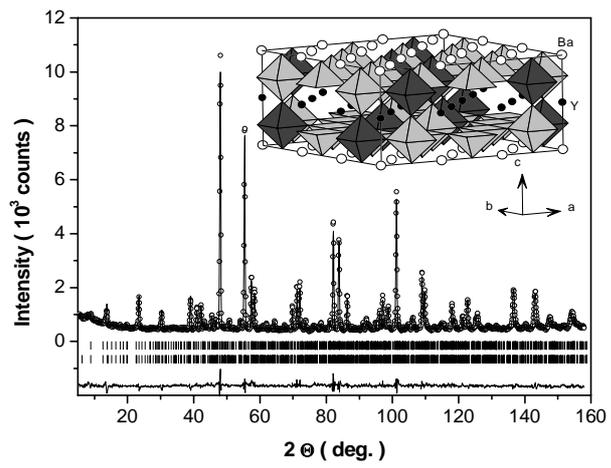

**Fig. 12**

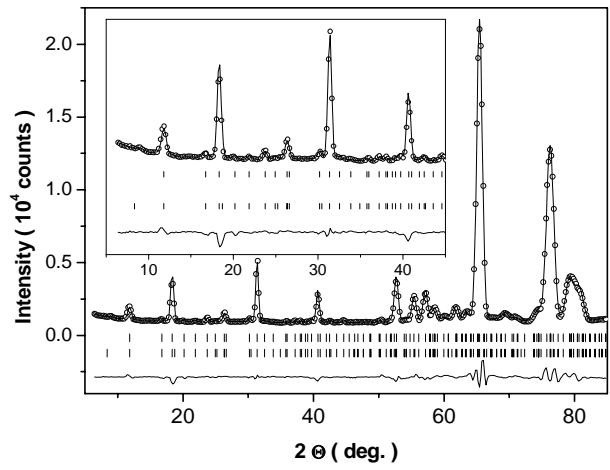

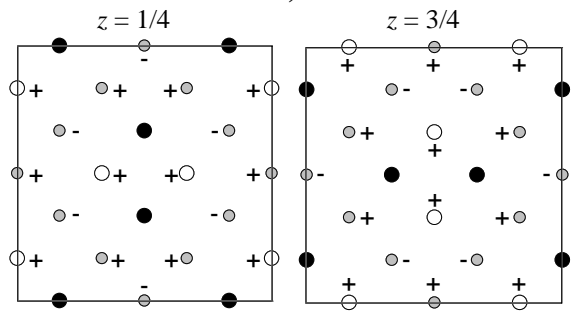

**Fig. 13**

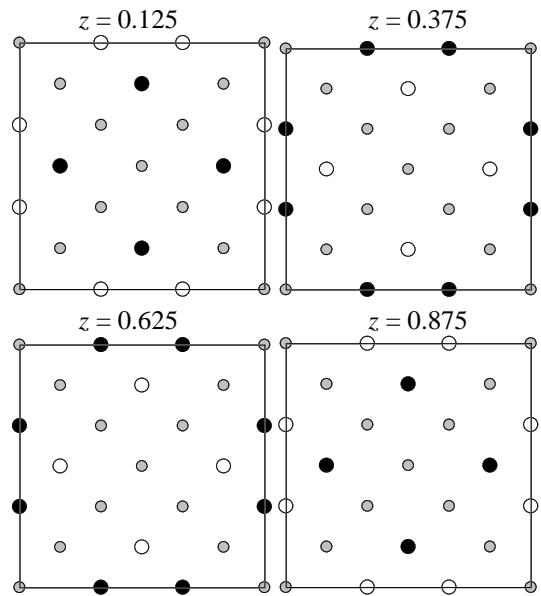

**a)** *I4/mmm* associated with $A_2^-/A_4^+$

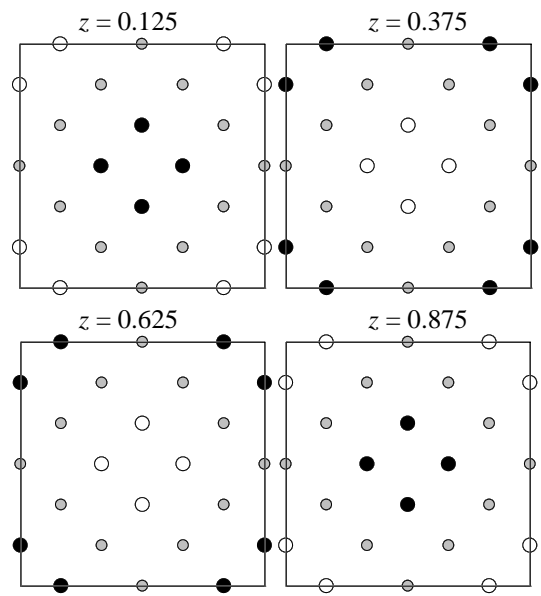

**b)** *I4/mmm* associated with $A_1^+/A_3^-$

**Fig. 14**

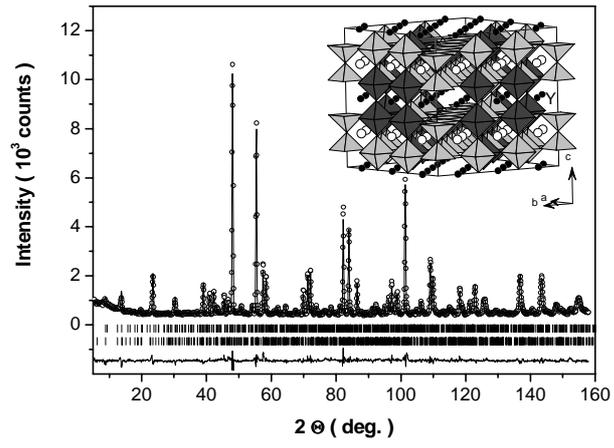

**Fig. 15**

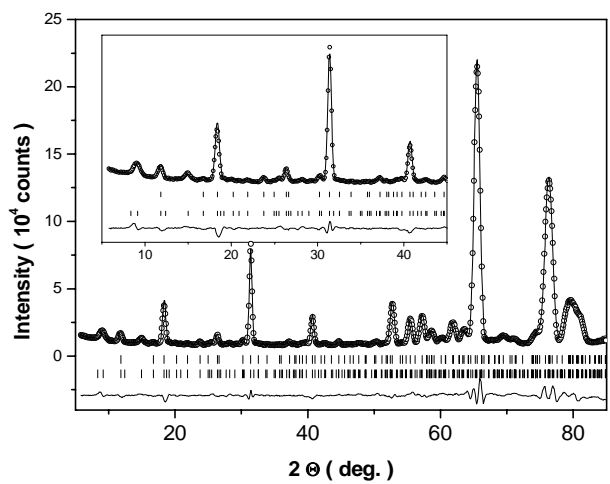

a)

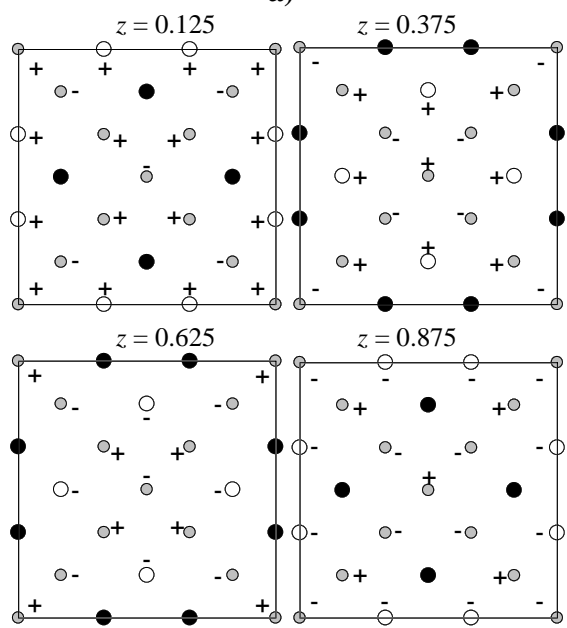

b)

**Fig. 16**